\documentclass[10pt,twocolumn]{IEEEtran}

\usepackage{graphicx}
\usepackage{latexsym}
\usepackage{amsfonts}
\usepackage{amssymb}
\usepackage{amsbsy}
\usepackage{amsmath}
\usepackage{multicol}
\usepackage{float}
\usepackage{subfigure}
\usepackage{algorithm}
\usepackage{algorithmic}
\usepackage[dvips]{epsfig}
\usepackage{dblfloatfix} 
\usepackage{color}
\usepackage{setspace}
\usepackage{fancyhdr}
\usepackage{epsfig}
\usepackage{threeparttable}
\usepackage{epsf,epsfig}
\usepackage{amsthm}
\usepackage[noadjust]{cite}
\usepackage{dsfont}
\usepackage{enumerate}
\usepackage{bbm}

\def\({\left(}
\def\){\right)}
\def\[{\left[}
\def\]{\right]}

\newtheorem{remark}{Remark}

\newtheorem{proposition}{Proposition}
\newtheorem{assumption}{Assumption}

\newtheorem{definition}{Definition}

\title{\LARGE Random Access with Opportunity Detection in Wireless Networks}
\author{\IEEEauthorblockN{Jinho Choi,
Seung-Woo Ko,
Koji Yamamoto,
and Seong-Lyun Kim}
\thanks{J. Choi and S.-L. Kim are with Yonsei University, Seoul, Korea (email: \{jhchoi, slkim\}@ramo.yonsei.ac.kr), 
S.-W. Ko is with the Division of Electronics and Electrical Information Engineering at Korea Maritime and Ocean University (KMOU), Busan, Korea
(email: swko@kmou.ac.kr),
K. Yamamoto is with Kyoto University, Kyoto, Japan (e-mail: kyamamot@i.kyoto-u.ac.jp).}}

\begin{document}

\bibliographystyle{plain}
\maketitle

\begin{abstract}
This letter proposes a novel random \emph{medium access control} (MAC) based on a transmission opportunity prediction, which can be measured in a form of a conditional success probability given transmitter-side interference. 
A transmission probability depends on the opportunity prediction, preventing indiscriminate transmissions and reducing excessive interference causing collisions. Using stochastic geometry, we derive a fixed-point equation to provide the optimal transmission probability maximizing a proportionally fair throughput. Its approximated solution is given in closed form. The proposed MAC is applicable to full-duplex networks, leading to significant throughput improvement by allowing more nodes to transmit.
\end{abstract}

\begin{IEEEkeywords}
Random MAC, transmission opportunity, proportional fairness, stochastic geometry \vspace{-10pt}
\end{IEEEkeywords}

\maketitle
\vspace{-5pt}
\section{Introduction} \label{sec:intro}
\vspace{-3pt}
A distributed \emph{medium access control} (MAC) plays a key role in efficient usage of wireless medium by allowing multiple users to share it without a central controller if they are temporally or spatially separated.
Due to its simplicity and practicality, it is preferred over a centralized MAC to facilitate the simultaneous control of massive number of devices, while frequent collisions and unfair resource allocation can happen in the same vein. 
To reduce the collision, a spectrum sensing approach is widely used where a \emph{transmitter} (TX) measures the interference before accessing the medium.   
The TX decides to access the medium if the interference is less than a predetermined threshold, e.g., \emph{Carrier Sensing Multiple Access} (CSMA) \cite{CSMA}. 
However, this deterministic decision may lead to inaccurate estimation of the transmission opportunity  
because the interference levels of a TX and the paired \emph{receiver} (RX) can be uncorrelated,
resulting in hidden- or exposed-node problems when the interference at the RX is respectively less or more than  that  at the TX \cite{Jiang08}. A straightforward approach to overcome the limitation is to add additional signaling protocols (see e.g., \cite{CMAP}, \cite{RTSS}), which are inappropriate in dense networks because it rather brings about excessive interference~\cite{Wang12}. 

This letter aims at developing a distributed MAC to maximize a network utility without additional signaling overhead. 
To this end, we propose a novel random access scheme based on the stochastic prediction of the transmission opportunity defined as a conditional success probability given a TX's measured interference. 
Contrary to the conventional random access without prediction (e.g.,  \cite{SARA}), it helps each node to access the medium only when the transmission is likely to be successful, leading to reducing the hidden- and exposed-node problems, and collisions. Besides, this MAC can control each node's transmission probability in a stochastic manner, which allows nodes to share the medium with achieving a proportional-fairness. 
Using a powerful tool of \emph{stochastic geometry} (SG) \cite{Haenggi12}, we derive a fixed-point equation to provide the optimal transmission probability in Proposition~\ref{prop1}. The approximated closed form solution is also given in Proposition~\ref{prop2}, giving useful insights for practical MAC design.    
It is worth noting that the proposed MAC is applicable to \emph{full-duplex} (FD) networks \cite{FD}. 
Compared with a \emph{half-duplex} (HD) based random MAC as in \cite{JMTWC}, 
it allows more nodes to access the medium, leading to significant throughput improvement.

\vspace{-11pt}

\section{System Model}
\vspace{-3pt}
Consider a Poisson bipolar network comprising two Poisson point processes (PPPs). 
A set of nodes $\left\{ {{\ell}} \right\}$ follows a homogeneous PPP ${\Phi _1}$ with density $\lambda$ and another set of nodes $\left\{ {{k}} \right\}$ also follows ${\Phi _2}$ where node $k$ is randomly located at a distance of $d$ from node $\ell$. 
Based on Displacement Theorem \cite{Haenggi12}, we regards $\Phi \equiv \Phi_1 \cup \Phi_2$ as a superposition of two homogeneous PPPs with intensity $\lambda$.
All nodes in the network are assumed to be FD-capable, and they can transmit and receive simultaneously.

Each node in the network transmits with its transmission probability specified in the sequel.
The signal between nodes $\ell$ and $k$ experiences Rayleigh fading with a unit mean $\left({h_{\ell k}}\sim\exp \left( 1 \right)\right)$. 
A \emph{signal-to-interference ratio} (SIR) of the transmitted signal from nodes $\ell$ to $k$ is expressed as: 
\vspace{-3pt}
\begin{equation} 
\mathsf{SIR}_k = \frac{{{P_\ell}{h_{\ell k}}{d^{ - \alpha }}}}{{\delta_k}{P_k}{\beta}+{\sum\limits_{m \in \Phi \backslash \left\{ \ell,k \right\}} {{\delta _m}{P_m}{h_{mk}}{{\left\| {m - k} \right\|}^{ - \alpha }}} }}, \vspace{-3pt}
\end{equation} 
\noindent where $P_m$ is the transmission power of node $m$. 
The parameter $\delta_m$ is an indicator that takes value $1$ if node $m$ transmits and $0$ otherwise, and $\alpha>2$ is a path-loss exponent. 
In the denominator, the first term represents the residual self-interference when both $\ell$ and $k$ are transmitting, i.e. FD communication, while the second one represents the aggregate interference from other TXs denoted by $\mathcal{I}_k$. The parameter $\beta \in [0,1]$ denotes residual self-interference of a node depending on a cancellation algorithm.
In this letter, we consider a unit transmission power $P_m=1$ for all $m\in \Phi$. 
The transmission is said to be successful when $\mathsf{SIR}_k \geq \theta$, where $\theta$ is an SIR threshold. That is to say, the desired signal cannot be decoded if SIR value is smaller than $\theta$.

\vspace{-11pt}

\section{Random Access}
\vspace{-3pt}
Every node determines the transmission probability by detecting transmission opportunity defined as follows.
\vspace{-5pt}
\begin{definition}
[Opportunistic Probability \cite{JMTWC}] \emph{The conditional success probability of the transmission from nodes $\ell$ to $k$ for a given TX-side interference $I$ is defined as
\vspace{-5pt}
\begin{equation} \label{eq:OP}
\mathsf{OP_{\ell}} \equiv \Pr \left[ {\left. {\mathsf{SIR}_{k} > \theta } \right| I} \right],
\vspace{-5pt} \end{equation} \normalsize
where $k$ denotes the dedicated RX of node $\ell$.}
\end{definition}\vspace{-5pt}
\noindent A typical node $\ell \in \Phi$ transmits with probability $p_\ell=\mathcal{F}\left( \mathsf{OP_\ell} \right)$ in the next slot and remains silent otherwise, where $\mathcal{F}\left( \mathsf{OP} \right)$ is a non-decreasing function. 
Although it is proved that a linear increasing $\mathcal{F}\left( \mathsf{OP} \right)$ gives a near-to-optimal performance to maximize an area spectral efficiency \cite{JMTWC}, it brings unfair resource allocation because the node with high OP monopolizes the radio resource. 
Hence, $\mathcal{F}\left( \mathsf{OP} \right)$ needs to be redesigned for a fair resource allocation. 
Before finding the optimal transmission probability, we describe why random access is preferred and how the nodes predict their OP in the following.

\vspace{-12pt}
\subsection{Advantages of Random Access} \label{sec:advan_RA}
\vspace{-3pt}

\begin{figure}
\centering 
{\includegraphics[width=9cm]{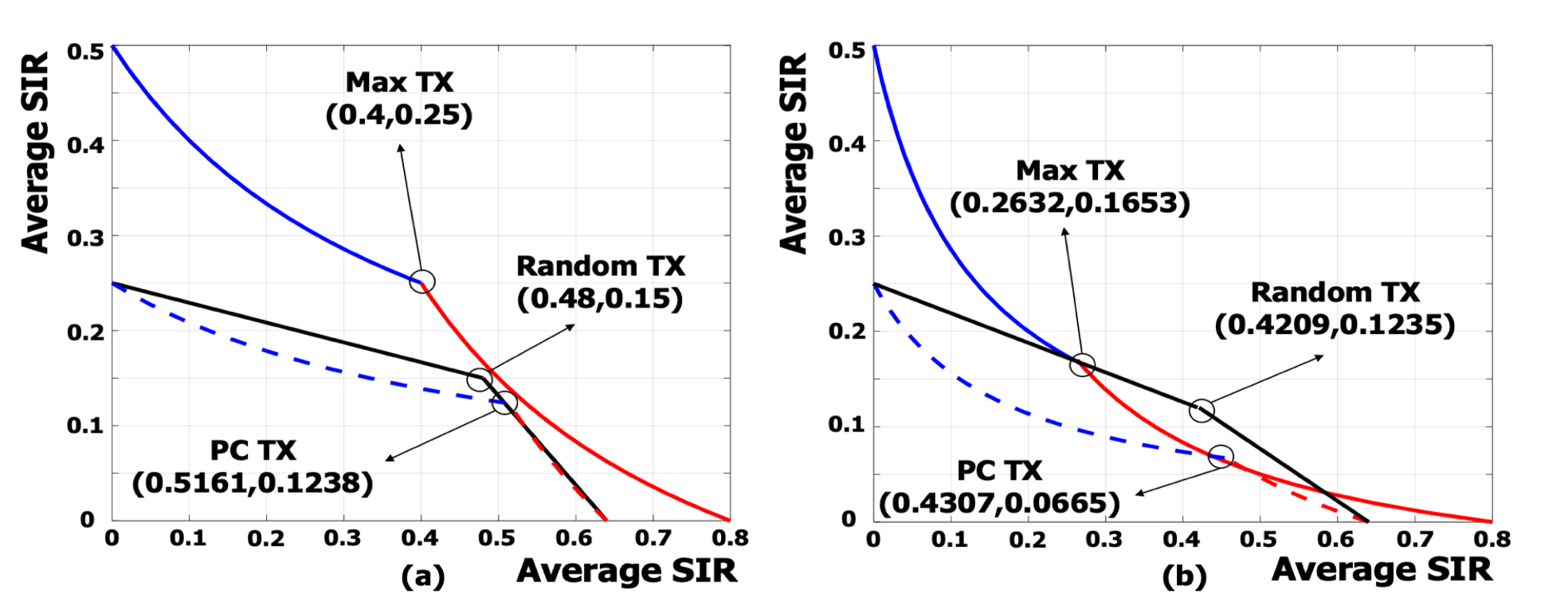}}
\caption{(a) SIR region for low mutual interference ($h_{11}=h_{22}=0.04, h_{12}=h_{21}=0.05$); (b) SIR region for high mutual interference ($h_{11}=h_{22}=0.04, h_{12}=h_{21}=0.15$).}\label{Fig:ProbvsPower}
\vspace{-15pt}
\end{figure}

We focus on two pairs in the network while the remaining interference from other nodes is constant. Figure \ref{Fig:ProbvsPower} shows the SIR regions of two RXs where the mutual interference conditions are small and large, respectively. We assume the OP values are $0.8$ (on X-axis) and $0.5$ (on Y-axis), and we refer to the node as high- and low OP nodes, respectively. 
Three distinguishable points denote SIR values of three access modes: i) \emph{maximal power transmission} (Max TX), ii) \emph{power-controlled transmission} (PC TX), iii) \emph{random access transmission} (Random TX).
The node reduces its transmission power by OP in PC TX, $P_{\ell}=\mathsf{OP_{\ell}}$, whereas the node sets its transmission probability, $p_{\ell}=\mathsf{OP_{\ell}}$ in Random TX while fixing $P_{\ell}$ to the maximal power.
In Figure \ref{Fig:ProbvsPower}a, there is small difference of 3\% in the sum of SIRs of three schemes while PC TX and Random TX spend less radio resources: \emph{(Tx time)} $\times$ \emph{(Tx power)}. 
It implies an increase in the transmission success rate and a decrease in loss of transmission opportunity caused by the exposed node problem. 
In Figure \ref{Fig:ProbvsPower}b, the SIR sums of Random TX and PC TX are even higher than Max TX as the mutual interference becomes large. 
The SIR sum as well as the SIR value of the low OP node are higher in Random TX than PC TX.
This means that Random TX achieves more equitable and efficient resource allocation as the network becomes interference-limited.
In PC TX, the low OP node experiences severe interference in every transmission, because the high OP node always transmits with a higher power. 
In Random TX, the low OP node has a chance to transmit with less interference when the high OP node is not transmitting.
This not only improves the fairness by increasing the average SIR of the low OP node, but also increases the SIR sum because the SIR increase of the low OP node is greater than the decrease of the high OP node. 
Based on what we have so far, we use Random TX as the basis of our MAC.

\vspace{-12pt}
\subsection{Prediction of Opportunity}
\vspace{-4pt}
A node measures the aggregate TX-side interference, $\mathcal{I}_\ell$, via the energy detection before the transmission.
A direct derivation of OP is intractable because the isotropic property is violated due to the common interferers of TX and RX. 
For the analytical tractability, we introduce the following assumption:
\begin{assumption}[Empty Ball] \emph{The interference $\mathcal{I}_\ell$ can be decomposed into the strongest- and the residual interferences. In an average sense, the strongest interference comes from the nearest interferer located on a circle defined as an empty ball. In other words, there is no interferer inside the ball. Furthermore, the other interferers are assumed to follow PPP outside the ball.}
\end{assumption}
\vspace{-5pt}
It is derived in \cite{JMTWC} that the expectation of the empty ball's radius for a given $I$ should satisfy the following condition: 
\vspace{-3pt}
\begin{equation} \label{eq:ri}
I{R}^\alpha-\frac{4\pi\lambda}{\alpha-2}R^2-1=0.
\end{equation} 
\noindent The above is solved in a closed form when $\alpha= 4$ as $R=I^{-\frac{1}{2}}\left[ \pi\lambda + \left( I + \pi^2\lambda^2  \right)^{\frac{1}{2}} \right]^{\frac{1}{2}}$. With the radius, the OP value is computed as in \cite{JMTWC}. \vspace{-10pt}

\section{Optimal Transmission Probability} \label{sec:prob}
\vspace{-3pt}
In this section, we derive the optimal transmission probability 
to maximize a proportionally fair throughput.
The expected throughput of node $\ell$ is $p_{\ell}q_k$, where 
$q_k$ denotes the success probability at its RX $k$. 
Then, we define the normalized link throughput as follows:
\vspace{-5pt}
\begin{definition}
[Normalized link throughput] \emph{The spatial average of the logarithmic link throughput is given by
\vspace{-4pt}
\begin{equation} \label{eq:logTP}
\mathsf{TP_{\ell}} = \mathop {\lim }\limits_{r \to \infty } \frac{1}{{\lambda \pi {r^2}}}\sum\limits_{k \in \Phi \cap b\left( r \right) } {\log \left( {{p_{\ell}}{q_k}} \right)}, \vspace{-4pt}
\end{equation}
where $b(r)$ denotes the circle with radius $r$ centered at the origin.}
\end{definition} \vspace{-5pt}

Using Palm distribution of the point process $\Phi$ and the ergodicity of the PPP, we can treat the nodes $\ell$ and $k$ as a typical pair located at the origin where the corresponding throughput of the pair, $\mathsf{TP}=\mathsf{TP_\ell} + \mathsf{TP}_k$. Then, we formulate a throughput-maximization problem of a typical FD pair to obtain the optimal transmission probability that achieves proportional fairness \cite{Baccelli14}: 
\vspace{-5pt}
\begin{eqnarray*} \label{ProblemForm}
\,\,\,\,\,\,\,\,\,\,\,\,\,\,\,\,\,\,\,\,\,\,\mathop {\max}\limits_{p_k} &&\mathsf{TP} = {\mathbb{E}_\ell}\left[ {\log \left( {{p_{\ell}}{q_k}} \right) + \log \left( {{p_k}{q_{\ell}}} \right)} \right] \,\,\,\,\,\,\,\,\,\,\,\,{\bf {(P)}}\\
{\rm{s}}{\rm{.t}}{\rm{.}}&&\,\,\,\,\,\,\,\,\,\,\,\,\,\,\,\,\,0 < {p_i} \le 1,\,\,\,\,i \in \Phi. 
\end{eqnarray*}
We note that {\bf (P)} is a general form of the network in which every node can transmit, i.e. FD network; we can also model a half-duplex network by setting $p_k=0$ for $k \in \Phi_2$. 
By solving {\bf (P)}, we obtain the probability as follows:
\vspace{-5pt}

\setcounter{equation}{12}
\begin{figure*}[b] \vspace{-12pt}
\hrulefill \vspace{-3pt}
\small
\begin{eqnarray} \label{coefficients}
\left\{ {\begin{array}{*{20}{c}}
\begin{array}{l}
\!\!\!\!\!\!{C_1} \!\!=\! \left( {10.4\! \times\! {\lambda ^{0.66}}\! + \!1.15} \right)\!\!\left( {3.5 \theta^{0.5}  \! + \!1.1} \right)\!\!\left( {0.5\! \times\! {d^2} \!- \!0.7} \right)\!\!\left( {4.5 \!\times \!{{10}^{ - 4}} \!\times\! {I^{ - 1.75}}\! - \!0.12} \right)\!\left( {-28.84 \!\times\! \exp\!\left({- \theta \beta {d^4}}\right)} \!+\!29.85\right)\\
\!\!\!\!\!\!{C_2} \!\!=\! 3.8 \!\times\! {10^{ - 4}}\! \times \!{I^{ - 1.75}}\! \times\! \left( {1.11\! \times \!{{10}^3} \!\times \!{\lambda ^2} \!+\! 3} \right)\!\left( {0.3 \theta^{0.5}  \! + \!2.3} \right)\!\left( {0.14\! \times \!{d^{2.5}}\! + \!6} \right)\!\left( {-0.129 \!\times\! \exp\!\left({- \theta \beta {d^4}}\right)} \!+\!1.129\right)\\
\!\!\!\!\!\!{C_3} \!\!= \! -\! \left( {{\lambda ^{1.63}}\! +\! 0.014} \right)\!\left( {0.7\! \times\! {\theta ^{ - 1}} \!+ \!0.5} \right)\!\left( {10 \!\times\! {d^{ - 4}} \!+\! 0.4} \right)\!\left( {0.9\! \times \!{I^{ - 1.763}} \!+ \!49} \right)\!\left( {0.02 \!\times\! \exp\!\left({- \theta \beta {d^4}}\right)} \!+\!0.98\right)
\end{array}&{\!\!\!\!\!\!\!\!\!\mathrm{if}\,\frac{{{{\left( {{{{R}} \mathord{\left/
 {\vphantom {{{R}} d}} \right.
 \kern-\nulldelimiterspace} d}} \right)}^2}}}{{\sqrt {{\theta  \mathord{\left/
 {\vphantom {\theta  2}} \right.
 \kern-\nulldelimiterspace} 2}} }} \le 10}\\
\begin{array}{l}
\!\!\!\!\!\!{C_1} \!\!=\! \left( {273 \!\times\! {\lambda ^{1.66}} \!+\! 2} \right)\!\left( {1.1{\theta ^{ - 1}} \!- \!5} \right)\!\left( {2.5 \!\times\! {d^{ - 4}} \!- \!0.7} \right)\!\left( {1.2\! \times \!{{10}^{ - 5}} \!\times\! {I^{ - 2.75}} \!- \!0.4} \right)\!\left( {7.96 \!\times\! \exp\!\left({- \theta \beta {d^4}}\right)} \!-\!6.958\right)\\
\!\!\!\!\!\!{C_2} \!\!=\! {10^{ - 4}} \!\times \!{I^{ - 2.75}} \!\times\! \left( {3.8 \!\times \!{{10}^3}\! \times \!{\lambda ^{2.66}} \!+ \!1.4} \right)\!\left( {2.7 \!\times\! {\theta ^{ - 1}}\! +\! 3.5} \right)\!\left( {34\! \times\! {d^{ - 4}}\! +\! 3} \right)\!\left( {0.5261 \!\times\! \exp\!\left({- \theta \beta {d^4}}\right)} \!+\!1.526\right)\\
\!\!\!\!\!\!{C_3} \!\!= \! - \!\left( {0.9 \!\times \!{\lambda ^{1.63}} \!+\! 0.013} \right)\!\left( {0.7 \!\times\! {\theta ^{ - 1}}\! + \!0.5} \right)\!\left( {17 \!\times\! {d^{ - 4}} \!+\! 0.7} \right)\!\left( {0.5\! \times\! {I^{ - 1.763}} \!+\! 30} \right)\!\left( {-0.046 \!\times\! \exp\!\left({- \theta \beta {d^4}}\right)} \!+\!1.051\right)\,\,\,\,\,\,
\end{array}&{\!\!\!\!\!\mathrm{o.w.}}
\end{array}} \!\!\!\!\!\right. .
\end{eqnarray}\normalsize
\end{figure*}

\begin{proposition} [Optimal transmission probability] \label{prop1}\emph{
The optimal transmission probability is given as $\min \left[ {p^*,1} \right]$ where $p^*$ is the solution of the following fixed-point equation:
\vspace{-5pt} \setcounter{equation}{4}
\begin{align} \label{FPE}
\frac{1}{{{p}}} =&-\!\frac{{{e}^{- \theta {d^\alpha } \beta}}-1}{\left({{e}^{- \theta {d^\alpha} \beta}} -1\right)p+1}+\frac{1}{{1 + \left( {{{R^\alpha} \mathord{\left/
 {\vphantom {{R^\alpha} {\theta {d^\alpha }}}} \right. \kern-\nulldelimiterspace} {\theta {d^\alpha }}}} \right) - {p}}}\nonumber\\ &+4\pi\lambda{d}^2 \int_{{R}/{d}}^{\infty} {\frac{s}{1-p+{s^\alpha}/{\theta}}ds}.
\end{align} 
Here, $R$ is the expected empty ball radius specified in \eqref{eq:ri}. When $\alpha=4$, the above equation is reduced as 
\begin{align} \label{FPE4}
\frac{1}{{{p}}} &=& -\frac{{{e}^{- \theta {d^\alpha } \beta}}-1}{\left({{e}^{- \theta {d^\alpha } \beta}} -1\right)p+1} + \frac{1}{{1 + \left( {{{R^\alpha } \mathord{\left/
 {\vphantom {{R^\alpha } {\theta {d^\alpha }}}} \right.
 \kern-\nulldelimiterspace} {\theta {d^\alpha }}}} \right) - {p}}} \nonumber \\ &&+ \frac{{2\pi \lambda {d^2}\sqrt \theta  }}{{\sqrt {1 - {p}} }}\left( {\mathrm{arctan}\left( {\frac{{{{\left( {{{{R}} \mathord{\left/
 {\vphantom {{{R}} d}} \right.
 \kern-\nulldelimiterspace} d}} \right)}^2}}}{{\sqrt {\theta \left( {1 - {p}} \right)} }}} \right)} \right).
\end{align} }

\vspace{-15pt}
\begin{proof}
First of all, we derive the success probability $q_k^{(\mathrm{H})}$ conditioned on $\Phi$ for a HD communication. \vspace{-10pt}

\small
\begin{align} 
&{\mathbb{E}_{{H_k}}}\left[ {\Pr \left( {\left. {\mathsf{SIR}_k \ge \theta } \right|{H_k},\Phi, \delta_k=0 } \right)} \right] \nonumber \\
&= \prod\limits_{i \in \Phi \backslash \left\{ {\ell,k} \right\}} { \left( {1 - \frac{{{p_i}}}{{1 + {{{{\left\| {i - k} \right\|}^\alpha }} \mathord{\left/
 {\vphantom {{{{\left\| {i - k} \right\|}^\alpha }} {\theta {d^\alpha }}}} \right.
 \kern-\nulldelimiterspace} {\theta {d^\alpha }}}}}} \right)} = {q_k^{(\mathrm{H})}}\label{eq:sucprob_HD}.
 \end{align}\normalsize 
\vspace{-10pt}
\noindent We also have the success probability $q_k^{(\mathrm{F})}$ as follows:

\small
\begin{equation} \label{eq:sucprob_FD}
{q_k^{(\mathrm{F})}}= \exp\left({- \theta {d^\alpha } \beta}\right)\prod\limits_{i \in \Phi \backslash \left\{ {\ell,k} \right\}} { \left( {1 - \frac{{{p_i}}}{{1 + {{{{\left\| {i - k} \right\|}^\alpha }} \mathord{\left/
 {\vphantom {{{{\left\| {i - k} \right\|}^\alpha }} {\theta {d^\alpha }}}} \right.
 \kern-\nulldelimiterspace} {\theta {d^\alpha }}}}}} \right)}.
\end{equation} \normalsize
The expected SIR of the node $k$, $q_k = \left( 1-p_k \right){q_k^{(\mathrm{H})}}+{p_k}{q_k^{(\mathrm{F})}}$.
Then, $\mathsf{TP}$ is represented as follows: \vspace{-10pt}

\begin{equation}
\begin{split}
\mathsf{TP}={\mathbb{E}_k}\!\left[ \log \left(p_\ell \left(1+p_k\left(\exp\left({- \theta {d^\alpha} \beta}\right)-1\right)\right){q_k^{(\mathrm{H})}}\right) \right. \\ 
\left. +\log \left(p_k \left(1+p_\ell\left(\exp\left({- \theta {d^\alpha} \beta}\right)-1\right)\right){q_\ell^{(\mathrm{H})}}\right) \right].
\end{split}
\end{equation}
The term $\log \left({q_k^{(\mathrm{H})}}\right)$ can be split into two terms depending on whether $i = z_m$ or not, where $z_k$ is the nearest interferer of node $k$, and we have the following equation by Campbell's Theorem. \vspace{-10pt}

\small 
\begin{equation}
{\mathbb{E}_k}\!\!\left[ {\sum\limits_{i \in \Phi \backslash \left\{ {\ell,k,{z_k}} \right\}} \!\!\!\!{\log \left( {1 - \frac{{{p_k}}}{{1 + {{{{\left\| {k - i} \right\|}^\alpha }} \mathord{\left/
 {\vphantom {{{{\left\| {k - i} \right\|}^\alpha }} {\theta {d^\alpha }}}} \right.
 \kern-\nulldelimiterspace} {\theta {d^\alpha }}}}}} \right)} } \right] \!=\! F\left( {{p_k},k} \right).
 \end{equation} \normalsize
\noindent Finally, we can maximize the expectation when the following derivative expression holds:
\vspace{-10pt}

\small 
\begin{equation} \label{eq:derieq}
\frac{1}{{{p_m}}} + \frac{{{e}^{- \theta {d^\alpha } \beta}}-1}{\left({{e}^{\!-\! \theta {d^\alpha\!} \beta}} \!-\!1\right)p_m\!\!+\!\!1}- \frac{1}{{1\! +\! \left( {{{R^\alpha \!} \mathord{\left/
 {\vphantom {{R^\alpha \!} {\theta {d^\alpha }}}} \right.
 \kern-\nulldelimiterspace} {\theta {d^\alpha }}}} \right) \!-\! {p_m}}} + \frac{{\partial F\left( {{p_m},m} \right)}}{{\partial {p_m}}} = 0,
\end{equation} \normalsize
where $m \in \left\{ {\ell,k} \right\}$ and 
\small
\begin{equation}
F\!\left( {{p_m},m} \right) \!=\! 2\lambda\! \int_{i \in {\mathbb{R}^2}\backslash b\left( {{R}} \right)}\! {\log \left(\! {1 \!\!-\!\! \frac{{{p_m}}}{{1 \!+\! {{{{\left\| {i - m} \right\|}^\alpha }}  {\theta {d^\alpha }}}}}} \right)di}.
\end{equation} \normalsize  \end{proof} 
\end{proposition}
\vspace{-10pt}
For a given measured interference, the optimal $p^*$ that maximizes the spatial average of the logarithmic throughput $\mathsf{TP}$ \eqref{eq:logTP} is determined by the fixed point equation \eqref{FPE}. 
It is clear that the RHS of the equation monotonically increases continually with respect to $p$ on $(0,1]$ because its differential value is positive while LHS monotonically decreases, thereby the equation has a unique solution. 
\vspace{-5pt}
\begin{remark} [Conservative Access] \label{remark:conservative}\emph{
Noting that each pair distance $d$ is different to each other in practice, our approach is said to be \emph{conservative} such that the resultant access probability in Proposition 1 provides a lower bound of the practical case. Specifically, \eqref{eq:sucprob_HD} represents a lower bound for the random distance with mean $d$ by Jensen's inequality. As a result, our approach is more likely to protect the transmissions of other devices, which is aligned with the mutual protection concept in spectrum-sharing networks \cite{Sharma18}.}
\end{remark}
\vspace{-10pt}
\begin{remark} [Effect of Self-interference Factor]\emph{
It is shown that as the residual self-interference factor $\beta$ decreases, the optimal transmission probability $p^*$ in Proposition \ref{prop1} increases. In other words, advanced self-interference cancellation technique makes it possible to reduce interference, allowing each node to access the medium more frequently. }
\end{remark}
\vspace{-5pt}

The nodes can numerically compute the optimal probability by fixed-point iteration based on \eqref{FPE}, but the probability calculations are frequent and time-sensitive.
To reduce the computational overhead, we approximate a closed-form expression of the optimal transmission probability as follows:


\vspace{-5pt}
\begin{proposition} [Approximated Transmission Probability]\label{prop2}\emph{
Given the path-loss exponent $\alpha=4$, the optimal transmission probability is given as $\min \left[ {p^*,1} \right]$ where $p^*$ is approximated as \vspace{-10pt} \setcounter{equation}{13}
 \begin{align}
p^*=\frac{-C_2+\sqrt{C_2^2-4C_1C_3}}{2C_1}, 
\end{align} 
\noindent where the coefficients $C_1$, $C_2$, and $C_3$ are given in \eqref{coefficients}.} \vspace{-5pt}
\begin{proof}
The $\mathrm{arctan}(x)$ rapidly increases when $x$ is relatively small, and the value is saturated as $x$ increases. 
We use two different functions, rational function in large interference and constant in small interference, for reflecting the distinct characteristics as follows: \vspace{-3pt}
\begin{eqnarray} \label{eq:arctan}
\arctan\! \left( x \right)\! \approx \! \left\{ {\begin{array}{*{20}{c}}
\!\!{\frac{1.632x-0.1037}{x+0.8967}}&{\!\! if\,x \le 10}\\
\!\!{\frac{\pi}{2}}&{\!\!o.w.}
\end{array}} \right. \vspace{-5pt}
\end{eqnarray}  \vspace{-1pt}
where $x = \frac{{{{\left( {{{{R}} \mathord{\left/
 {\vphantom {{{R}} d}} \right.
 \kern-\nulldelimiterspace} d}} \right)}^2}}}{{\sqrt {{\theta  \mathord{\left/
 {\vphantom {\theta  2}} \right.
 \kern-\nulldelimiterspace} 2}} }}$. The root mean square error between the function and $\arctan(x)$ is $0.018$ on $x\in[0,10]$.

Then, we can approximate $p^*$ as a solution of the following quadratic equation by curve fitting. 
\vspace{-5pt} 
\begin{equation}
C_1\cdot p^2 + C_2 \cdot {p} +C_3 = 0. \vspace{-3pt}
\end{equation} 
The coefficients for each section are given by \eqref{coefficients}.
%
\end{proof}
\end{proposition}
\vspace{-10pt}
\begin{remark} [Effect of Parameters]
\emph{Proposition \ref{prop2} shows that the approximated transmission probability decreases as node density $\lambda$, pair distance $d$, SIR threshold $\theta$, and measured interference $I$ increase. 
For large interference regime, two parameters $\theta$ and $d$ have large influence on the probability because the effect of the desired signal's attenuation is large.
On the other hand, the influence of $\lambda$ becomes large in small interference regime because the growth of interference due to the increase of $\lambda$ greatly reduces the success probability.}
\end{remark}
\vspace{-10pt}
\section{Numerical Results} \label{sec:numerical}
\vspace{-3pt}
\begin{figure}
\centering 
{\includegraphics[width=9cm]{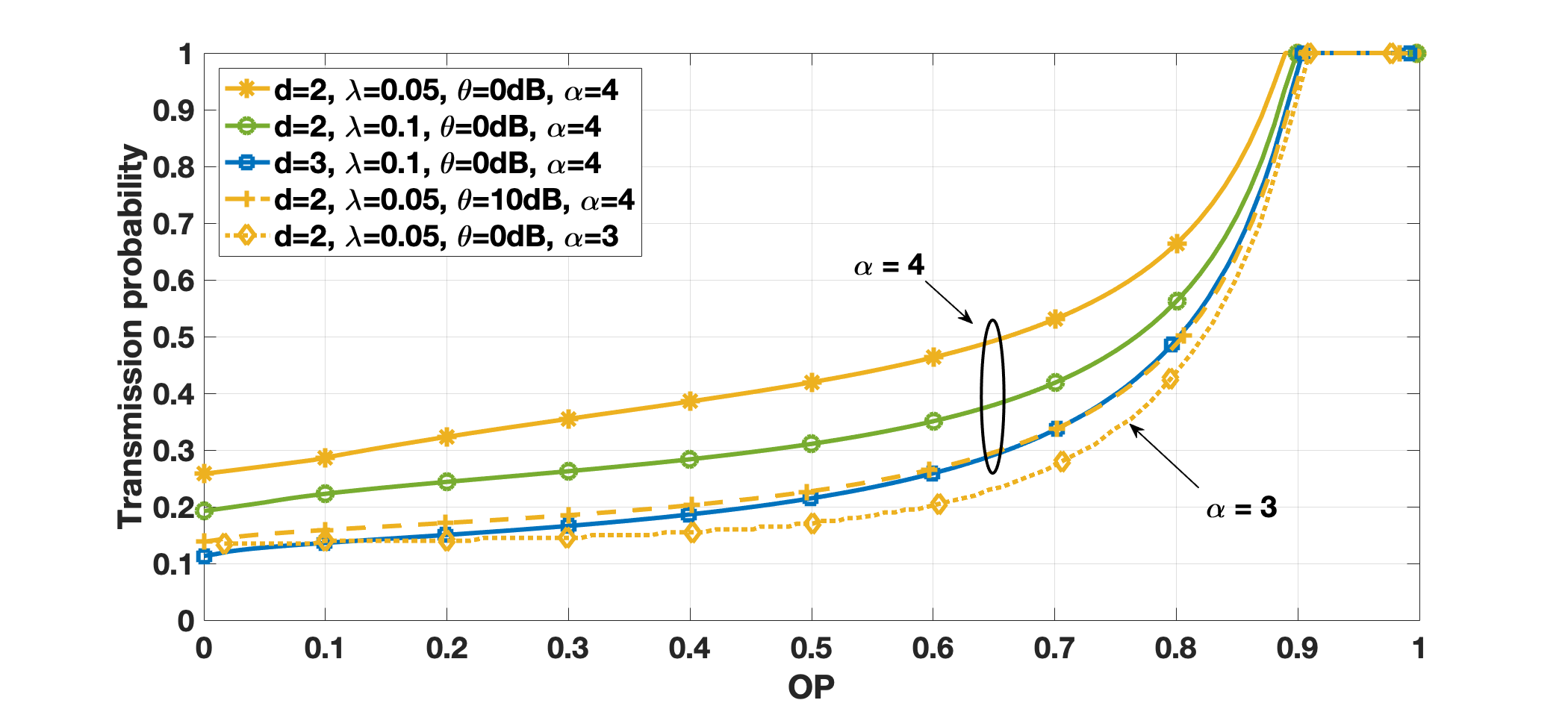}}
\caption{Optimal transmission probability as a function of OP ($\beta=-110dB$).}\label{Fig:OptProb} \vspace{-10pt}
\end{figure}

Figure \ref{Fig:OptProb} shows the optimal transmission probability obtained by Proposition \ref{prop1} as a function of OP. 
Based on \cite{FD}, we set the residual self-interference factor $\beta=-110dB$. The effect of $\beta$ cannot be shown due to the lack of space, but the result remains unchanged if $\beta$ is less than $-50dB$.
We figure out that the probability increases with a quasi-convex shape as OP increases. 
A transmission should be restricted if its success probability is low, since the communication failure causes unnecessary interference to the network.
As density and threshold increase, the node transmits more conservatively even with the same OP in order to generate less interference.
In a similar vein, the smaller the parameter $\alpha$, the lower the probability of transmission even with the same OP, because the transmission of a node is more interfering with the others.
Nevertheless, the nodes have non-zero probability even if OP is zero as a result of proportionally fair resource allocation. 

\begin{figure}
\centering 
{\includegraphics[width=9cm]{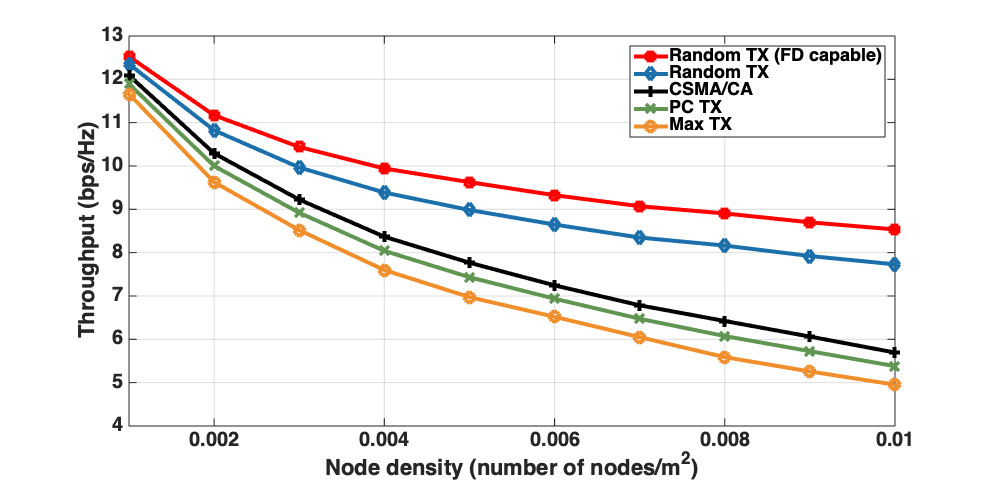}}
\caption{Throughput according to the access schemes ($\alpha=4$, $\theta=3dB$, $d=3m$, and $\beta=-110dB$)}\label{Fig:TP} \vspace{-15pt}
\end{figure} 

Figure \ref{Fig:TP} shows the link throughput according to the different access schemes: Max TX, PC TX, Random TX and CSMA/CA. The carrier sensing threshold in CSMA/CA and the transmission power of nodes are set as $-30$ dBm and $23$ dBm, respectively. 
The throughput difference between Random TX and CSMA/CA becomes larger as the node density increases, because applying OP enables nodes to determine a suitable probability even in the interference-limited situation while they wastes radio resources for the contention in CSMA/CA. 
Although PC TX also controls the transmission power by OP, the gain is insignificant because the weakened desired signal cancels out the effect of reduced interference. 
Random TX prevents excessive interference and exploits spectrum reusability in FD networks as well, resulting in an average throughput gain of $12$\%.

\begin{figure}[t]
\centering 
\subfigure[]{
{\includegraphics[width=8.5cm]{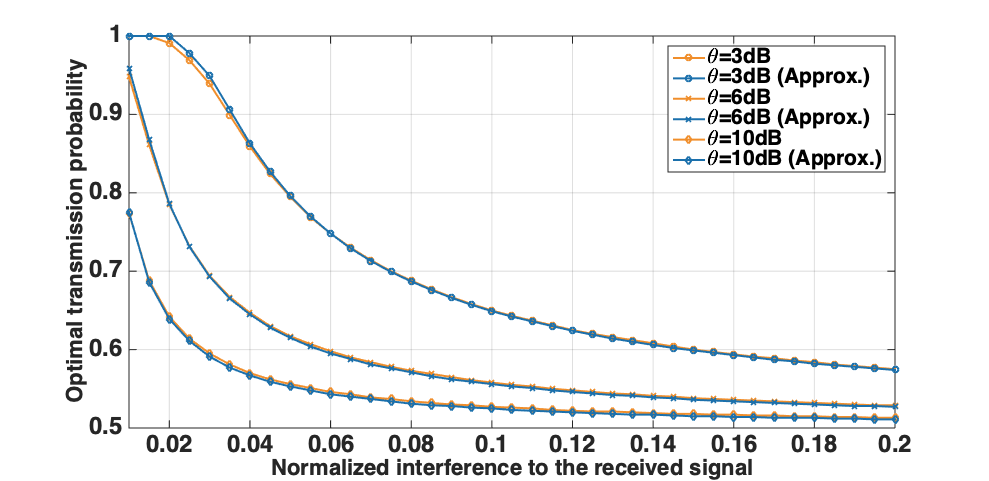}}
}
\subfigure[]{
{\includegraphics[width=8.5cm]{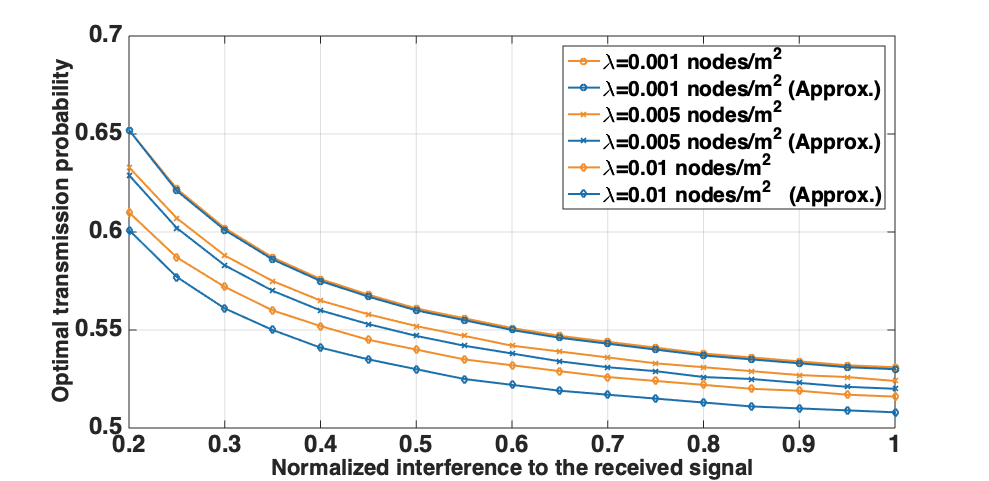}}
}
\caption{(a) Optimal transmission probabilities and its approximation as $\theta$ varies when $\alpha=4$, $d=2 m$, $\lambda=0.001$ nodes/$\text{m}^2$; (b) Optimal transmission probabilities and its approximation as $\lambda$ varies when $\alpha=4$, $d=2$, $\theta=1$. }\label{Fig:Approx}
\vspace{-15pt}
\end{figure}

Figure \ref{Fig:Approx} shows the optimal transmission probabilities and its approximation for a verification of the approximation used in Proposition \ref{prop2}. The approximation is accurate in the fact that the averaged and maximum errors are 0.5\% and 1.4\% in small interference environment (Figure \ref{Fig:Approx}a), 1.76\% and 1.99\% in large interference region (Figure \ref{Fig:Approx}b). 
\vspace{-5pt}
\section{Conclusion}
\vspace{-3pt}
We proposed a random access scheme that prevents indiscreet transmissions based on the prediction of transmission opportunity in wireless networks.
We also provided a fixed-point equation and an approximated form to find the optimal transmission probability that maximizes proportionally fair throughput.
The proposed scheme enables each transmission node to control its transmission probability, thereby transmission pairs can control their duplexing mode flexibly if they have FD capability.


\vspace{-15pt}
\begin{thebibliography}{1}


%

\bibitem{CSMA}
I. W. Group {\em et al.}, 
\newblock IEEE 802.11-2007: Wireless LAN Medium Access Control (MAC) and Physical Layer (PHY) Specifications,
\newblock 2007.

\bibitem{Jiang08}
L. B. Jiang and S.C. Liew,
\newblock ``Improving throughput and fairness by reducing exposed and hidden nodes in 802.11 networks,"
\newblock {\em IEEE Transactions on Mobile Computing}, 
\newblock vol. 7, no. 1, pp. 34-49,
\newblock 2008.


\bibitem{CMAP}
M. Vutukuru, K. Jamieson and H. Balakrishnan, 
\newblock ``Harnessing exposed terminals in wireless networks,"
\newblock in {\em Proceedings of USENIX Symposium on Networked Systems Design and Implementation,}
\newblock 2008. 

\bibitem{RTSS}
K. Mittal and E. Belding,
\newblock ``RTSS/CTSS: Mitigation of exposed terminals in static 802.11-based mesh networks,"
\newblock in {\em Proceedings of IEEE Workshop on Wireless Mesh Networks,}
\newblock 2006. 

\bibitem{Wang12}
L. Wang, K. Wu and M. Hamdi,
\newblock ``Combating hidden and exposed terminal problems in wireless networks,"
\newblock {\em IEEE Transactions on Wireless Communications,}
\newblock vol. 11, no. 11, pp. 4204-4213,
\newblock Oct.
\newblock 2012.  



\bibitem{SARA}
D. M. Kim and S. -L. Kim, 
\newblock ``Exploiting regional differences: A spatially adaptive random access,"
\newblock {\em IEEE Transactions on Wireless Communications,}
\newblock vol. 14, no. 8, pp. 4342-4352, 
\newblock Aug.
\newblock 2015.





\bibitem{Haenggi12}
M. Haenggi,
\newblock {\em Stochastic Geometry for Wireless Networks.},
\newblock Cambridge University Press,
\newblock 2012.

\bibitem{FD}
D. Bharadia, E. McMilin, and S. Katti,
\newblock ``Full duplex radios,"
\newblock in {\em Proceedings of ACM SIGCOMM},
\newblock Aug.
\newblock 2013.

\bibitem{JMTWC}
J. Kim, S.-W. Ko, H. Cha, S. Kim, and S.-L. Kim, 
\newblock ``Sense-and-predict: Harnessing spatial interference correlation for opportunistic access in cognitive radio networks," 
\newblock arXiv:1802.01088 [cs.IT],
\newblock 2018.


\bibitem{Baccelli14}
F. Baccelli, B. Blaszczyszyn, and C. Singh,
\newblock ``Analysis of a proportionally fair and locally adaptive spatial aloha in poisson networks,"
\newblock in {\em Proceedings of IEEE INFOCOM,}
\newblock 2014.

\bibitem{Sharma18}
S. K. Sharma et al., ``Dynamic spectrum sharing in 5G wireless networks with full-duplex technology: Recent advances and research challenges,"
\newblock {\em IEEE Communications Surveys \& Tutorials},
\newblock vol. 20, no. 1, pp.674-707,
\newblock first quarter,
\newblock 2018. 

\end{thebibliography}
\end{document}